\documentclass{llncs}
\pdfoutput=1

\usepackage[utf8]{inputenc}
\usepackage[hidelinks]{hyperref}
\usepackage{cite}
\usepackage{graphicx}
\usepackage{amsmath}
\usepackage{esvect}

\author{
	Ivica Obadić
	\and
	Gjorgji Madjarov 
	\and
	Ivica Dimitrovski
	\and
	Dejan Gjorgjevikj
}

\institute{
	Faculty of Computer Science and Engineering, Ss. Cyril and Methodius University,
	Rugjer Boshkovikj 16, 1000 Skopje, Macedonia \\ 
	\email{
		\href{mailto:obadic.ivica@gmail.com}{obadic.ivica@gmail.com},
		\href{mailto:gjorgji.madjarov@finki.ukim.mk}{gjorgji.madjarov@finki.ukim.mk}
		\href{mailto:ivica.dimitrovki@finki.ukim.mk}{ivica.dimitrovki@finki.ukim.mk}
		\href{mailto:dejan.gjorgjevikj@finki.ukim.mk}{dejan.gjorgjevikj@finki.ukim.mk}
		}
}

\title{Addressing Item-Cold Start Problem in Recommendation Systems using Model Based Approach and Deep Learning}

\begin{document}

	\maketitle

	\begin{abstract}
		Traditional recommendation systems rely on past usage data in order to generate new recommendations. Those approaches fail to generate sensible recommendations for new users and items into the system due to missing information about their past interactions. In this paper, we propose a solution for successfully addressing item-cold start problem which uses model-based approach and recent advances in deep learning. In particular, we use latent factor model for recommendation, and predict the latent factors from item's descriptions using convolutional neural network when they cannot be obtained from usage data. Latent factors obtained by applying matrix factorization to the available usage data are used as ground truth to train the convolutional neural network. To create latent factor representations for the new items, the convolutional neural network uses their textual description.	The results from the experiments reveal that the proposed approach significantly outperforms several baseline estimators. 
	\end{abstract}
	
	\section{Introduction}
		
		Advances of internet and popularity of modern interactive web applications led to creation of massive amounts of data by internet users. This data becomes useful only when it is analyzed and turned out into valuable information that can be used in future.	\textit{Big data} emerged as a concept which describes the challenge of smart management of this vast amounts of data. Common for almost all \textit{big data} challenges is that no explicit algorithms that can solve such problems in finite number of steps have yet been discovered. For example, there is not explicit algorithm which can correctly predict customers future behavior from past data. Therefore, in order to tackle \textit{big data} challenges and needs, \textit{artificial intelligence} and it's fields, especially \textit{machine learning} and \textit{data mining} became actual and extremely popular.
		\par
		The users in 'big data world' are faced with huge amount of information and choice of large sets of items. Only a small subset of those items fit to the interests of the users. Therefore, to attract users and keep their attention, the applications must provide a \textbf{personalized} content to the user. Personalized content reduces the time of search for the users, prunes the large search space and directs users to the content of their interest. These needs brought recommendation systems live on the scene.
		\par
		Recommendation systems apply machine learning and data mining techniques over past historical data about users interactions with items in the system. They create a model for the users and the items and then generate new recommendations on basis of that model. There are two major approaches for building recommendation systems (achieving this goal) denoted as collaborative filtering approaches and content-based approaches.
		
		\subsection{Collaborative Filtering}
		Collaborative filtering (CF) methods produce user specific recommendations of items based on ratings or usage without any additional information about either item or users \cite{Rajaraman:2011:MMD:2124405}. There are two main types of collaborative filtering approaches \cite{Ricci:2010:RSH:1941884}:
		\begin{itemize}
			\item \textit{Neighborhood-based} methods recommend items based on relationships between items or, alternatively, between users.
			\begin{itemize}
			\item \textit{User-user based}  create recommendations based on similarity between users. Similarity between any two users most often is determined on the basis on similarity of expressed preferences given by both users for same set of items \cite{Segaran:2007:PCI:1406354}.
			\item \textit{Item-item based}  approaches use known ratings made by the same user on similar items. They recommend items to a user that are similar with the items for which user has expressed positive preference in the past. This approaches offer better scalability and improved accuracy \cite{bell2007scalable}. 
			\end{itemize}
			\item \textit{Model-based} methods, such as matrix factorization (aka, SVD, SVD++, Time SVD++) \cite{Ricci:2010:RSH:1941884} \cite{Koren:2008:FMN:1401890.1401944} \cite{Koren:2009:CFT:1557019.1557072} \cite{Oord:2013:DCM:2999792.2999907} model latent characteristics of the users and items. These methods transform items and users to the same latent factor space that describes the users and items by inferring user feedback. These methods become very popular since their use for movie recommendation in the Netflix Prize \cite{Bennett07thenetflix}. 
		\end{itemize}
				
		\subsection{Content-based approach}
		Users and items in recommendation systems that use content-based approaches are modeled using profiles \cite{Rajaraman:2011:MMD:2124405} that represent their most significant characteristics and features. Profiles are usually comprised of binary or real values and in literature are often called \textit{feature vectors}. Item profiles can be constructed by using their metadata. For example if items are movies, then the director and the actors of certain movie can be used as components for the items feature vector. In most cases item metadata by itself does not provide enough information to create rich model that generates sensible recommendations. Therefore, additional features are created by analyzing item content and other data provided for the item such as expressed preferences for that item by the users in the system. In the work presented in this paper we are relying on such features by learning \textit{latent factor} models for users and items in the system.
		
		\subsection{Cold-Start problem}
		The rapid growth of users and content in the applications causes recommendation systems to face common and constant issue of generating recommendations for new users and/or new items that appear in the system. This problem in the literature is known as \textit{cold start} problem \cite{DBLP:journals/corr/BernardiKKM15}, and means that the systems do not have enough data to generate personalized recommendation for a new user (that has just entered the system and has no previous history), or for a new item (product) added to the system. Due to the massive amount of content generated each day in the applications, recommendation systems especially struggle to properly recommend the new items. These two problems are known as a user-cold start and item-cold start problem.
		\par
		Therefore, the goal of this paper is to propose a successful approach for addressing item-cold start problem in recommendation systems. Here, we propose to use a latent factor model for recommendation, and predict the latent factors from item's descriptions when they cannot be obtained from usage data. In particular, our approach consists of learning latent factors for a new item from its reviews by using deep convolutional neural network first and then use those latent factors for rating the new item to each user by using model-based approaches (\textit{SVD++}). We evaluate our approach on a dataset provided by Yelp in 'RecSys2013: Yelp Business Rating Prediction' challenge. The results are compared with multiple baseline estimators.

		\subsection{Organization}
		The rest of the paper is organized as follows. Section 2 gives description of the dataset used in experiments of this paper and explains it's suitability for the topic of this research.	Section 3 describes the algorithm used for learning item latent factors from item review. Special attention is given to the architecture of the convolutional neural network.	Experiments and results are presented and discussed in section 4. In the last section (Section 5), we make the conclusions and discuss possible further directions for extending our work.

	\section{The Dataset}	
	\begin{figure}
		\begin{minipage}[b]{0.5\textwidth}
			\includegraphics[scale=0.075]{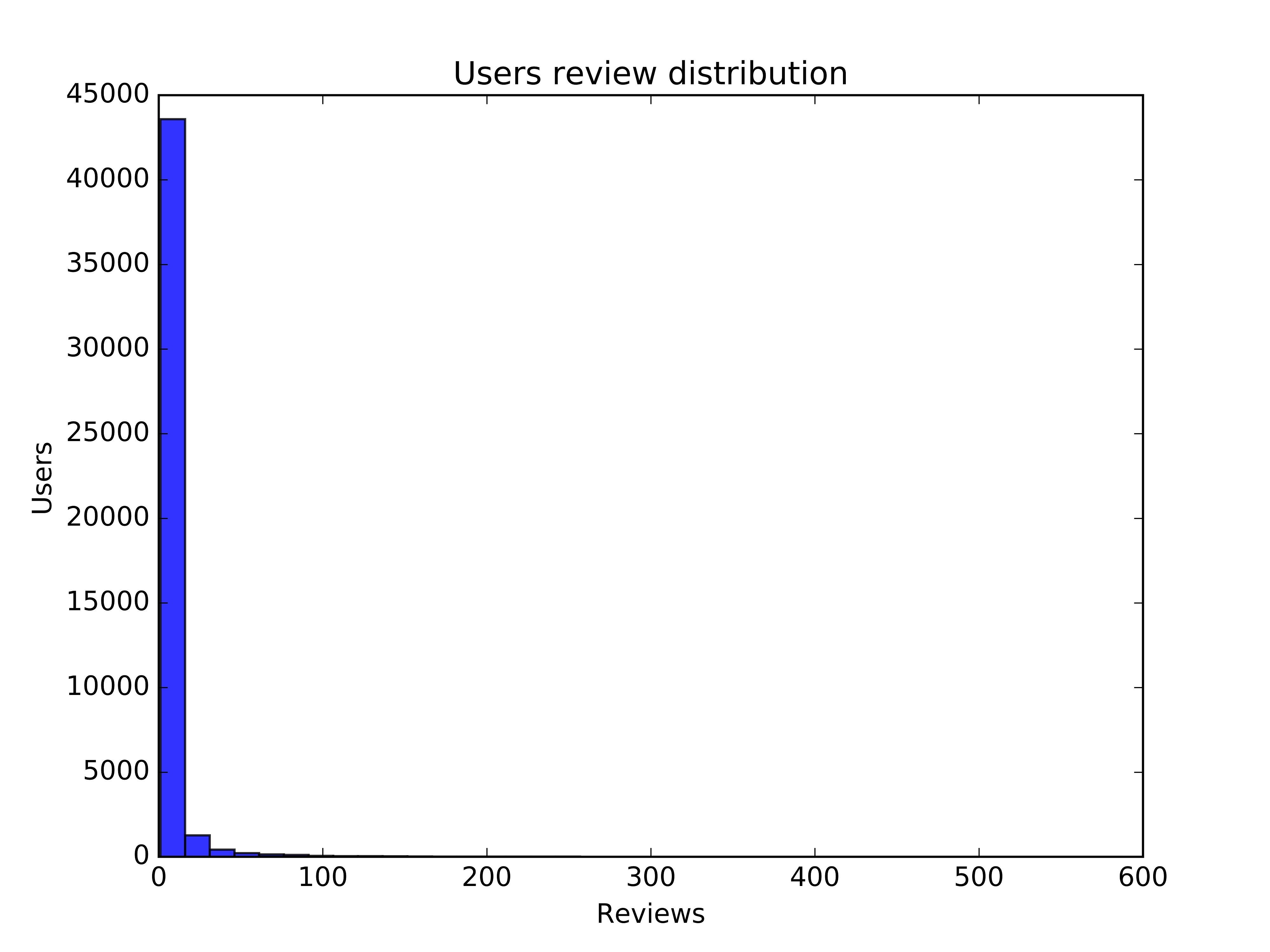}
			\caption{Users review distribution}
			\label{fig:users_review_distribution}
		\end{minipage}
		\begin{minipage}[b]{0.5\textwidth}
			\includegraphics[scale=0.075]{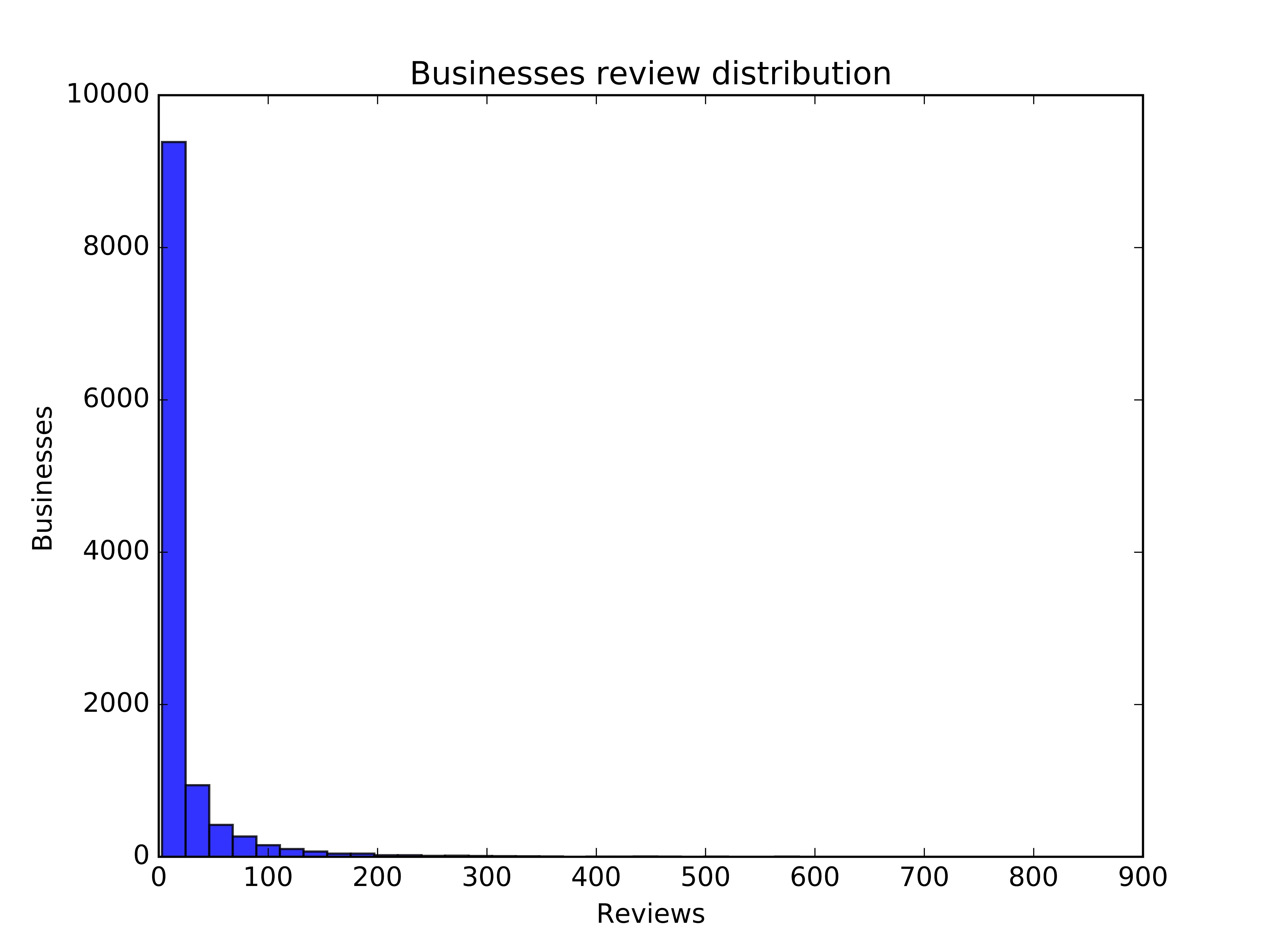}
			\caption{Businesses review distribution}
			\label{fig:businesses_review_distribution}
		\end{minipage}
	\end{figure}
	
	\textbf{Yelp} dataset\footnote{https://www.kaggle.com/c/yelp-recsys-2013/data} used in the experiments of this paper provides real-world data about users, businesses, reviews and checkins for Phoenix, AZ metropolitan area. Of interest of our research are the reviews of the businesses in the dataset. There are total \textit{229 907} reviews given by \textit{45 981} distinct users for \textit{11 537} distinct businesses. 	This dataset properly simulates the cold start problem because there are known only \textit{0.00043}\% of all possible users preferences for businesses.	
	Fig. \ref{fig:users_review_distribution} and fig. \ref{fig:businesses_review_distribution} show the user review distribution and the distribution of reviews for businesses in the dataset respectively. From this figures can be noticed that each user rated only few businesses that leads to a very small number of reviews for majority of businesses in the dataset.
	
	\section{Predicting Latent Factors from Item Description}
	
	First step in our proposed solution consists of learning latent factors with matrix factorization approach (SVD++ in our case). We represented each user and business with 20 latent factors. Models learned solely with SVD++ are unable to generate predictions in item-cold start problem scenario. Main reason is that latent factors of the new business in the system can not be computed because there are no existing preferences in the database for those businesses. 
	Hence, instead from user preferences data, latent factors for a new business could be learned from other data available for that business. In Yelp dataset, each business has text which represents user textual description about review. Predicting latent factors for a given business from its description is a regression problem. Any traditional regression technique can be used for mapping the feature representation of a business description to latent factors. In our work we use a deep convolutional neural network for mapping the business descriptions directly  to latent factors without a need of separate feature extraction approach. 
	Latent factor vectors obtained by applying SVD++ to the available usage data are used as ground truth to train the prediction models. Since there can be multiple reviews for same business, as description for each business, we took the text from the review that has most votes from the other users.
	
	\subsection{Deep Convolutional Neural Network}
	Convolutional neural networks are firstly used in areas of computer vision and speech recognition.
	Deep learning approaches using convolutional neural networks became extremely popular after the high achievments by using this approach in the Large Scale Visual Recognition Challenge\footnote{http://www.image-net.org/challenges/LSVRC} contest in the area of image classification \cite{NIPS2012_4824}. 
     In \cite{Collobert:2011:NLP:1953048.2078186}, \cite{DBLP:journals/corr/KalchbrennerGB14} and \cite{DBLP:journals/corr/Kim14f} are given examples on usage of convolutional neural network architectures in the field of Natural Language Processing.
	In our work, we used a convolutional neural network for learning a function that directly maps the business reviews to latent factor vectors representation.	
	
	\subsubsection{Architecture}
	The architecture of the convolutional neural network used in our research is similar to architectures proposed in \cite{Collobert:2011:NLP:1953048.2078186}, \cite{DBLP:journals/corr/Kim14f} and \cite{stojanovski2015twitter}.
	Our network is composed of the following four layers: input (embedding) layer, convolutional layer, max-pooling layer and the final output layer.
	\par
	The system works by first, constructing a lookup table, where each word is mapped to an appropriate feature vector or word embedding. One way of building the lookup table is by randomly initializing the word vectors. However, previous work showcased that initializing the word embeddings with pre-trained ones provides for better performance as opposed to random initialization. In this work, the input layer of the network maps the business description to a matrix of 300 dimensional Glove word embeddings \cite{pennington2014glove}. Each word in the description is mapped to it's corresponding vector and those vectors are concatenated into a matrix. Alternatively, we can consider our input layer as a function that does the following:
	\begin{equation*}
		f(word_1, word_2, ...,word_n) \mapsto 
		\begin{pmatrix}
  		\vv{word_1}\\
 		\vv{word_2}\\
  		\vdots\\
  		\vv{word_n} 
 	\end{pmatrix}
	\end{equation*}	
	Word vectors are additionally updated with backpropagation during the training of the network in order to adapt to the meaning of the corpus in the dataset.
	\par
	The convolution operation is then applied to the mapped word embeddings and max-over-time pooling in order to get a fixed sized vector. In this step, we apply 50 filters, We use rectified linear units (ReLU) \cite{icml2010_NairH10} in the convolutional layer and sliding window of size 4.
	By applying the convolution operation with this setup, input matrix of shape N x 300, where N represents number of words in business description, is transformed into output matrix of shape N x 50.
	The max-pooling layer subsamples the output of convolutional layer by using region size of N x 1. 
	With this size of region, input to this layer is transformed into a 50 neurons which correspond to the highest values computed by each filter.
	Neurons from the the max-pooling layer are fully connected to the 20 neurons of the output layer which represent the latent factors for the businesses. 
	
	\subsection{Pre-Processing Descriptions}
	Pre-processing businesses descriptions is necessary step in order to fed the descriptions as an input to the neural network.
	At first, each description is padded with zeros on the end, so that the length of each description in dataset is equal to the length of the longest description. Then, each word is replaced with it's corresponding Glove word embedding. For the words that were not found in the set of Glove words, the \textit{edit distance} \cite{cormode2007string}  between the corresponding word and each of the words in the Glove set was computed and the word is replaced with the closest word from the glove set if the edit distance is less than or equal to 2, or with a value randomly initialized in the interval [-0.25, 0.25] if no such word exists in the glove set. Random initialization of word embeddings is approach suggested by Kim in \cite{DBLP:journals/corr/Kim14f}.

	\section{Experiments}

	\subsection{Experimental setup}
	
	For our experiments, we took 80\% of the businesses and reviews for training, and the other 20\% for testing the proposed approach.
	In order to properly simulate the item-cold start problem, the split that we performed also satisfies the following constraint: each business that occurs in the reviews in the test set, does not occur in any of the reviews in the training set. 
	\par
	The distribution of reviews for businesses in dataset (shown in fig. \ref{fig:businesses_review_distribution}) shows high imbalance in the number of reviews per business and requires careful decision about the split.
	Large number of reviews for a business in general allows more accurate and broader description of the business and reduces the outliers effect when training the model. On the other hand, a model learned from businesses with only few reviews in the dataset is much more prone to outliers and user bias effects.
	In order to test the predictive performance of our model on the two types of businesses according to the number of given reviews, we wanted our split to support two separate evaluations of generated predictions.
	Separate evaluations need to be performed on predictions for businesses with small number of reviews and for businesses with large number of reviews.
	Taking this into account, we define two different test sets:
	\begin{itemize}
		\item \textbf{Test set 1} consists of reviews from the 15\% of businesses with smallest number of reviews in the dataset, which have at least one review with 5 or more votes from other users. This set contains 5 625 review samples in total.
		\item \textbf{Test set 2} consists of reviews from the second half (5\%) of the top 10\% reviewed businesses which also have at least one review with 5 or more votes from other users. This set contains 45 582 review samples in total.
	\end{itemize}
	The training set contains the rest 178 424 reviews (80\%).
	All further experiments described in following subsections are performed over these training set and test sets.
	
	\subsection{Parameter instantiation}
\begin{figure}
				\includegraphics[width=0.9\textwidth, scale=0.0001]{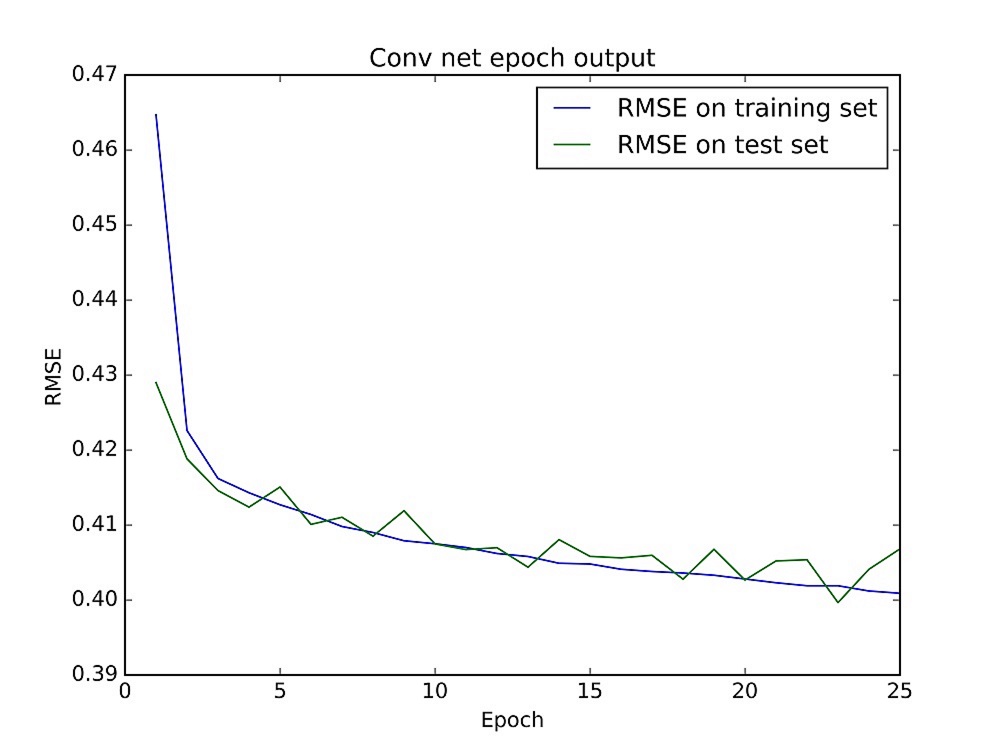}
				\caption{Convolutional neural network training}
				\label{fig:conv_net_train_test_error}
			\end{figure}

	In order to properly rate the businesses for each user, we proceed according to the following steps: 
	\begin{enumerate}
	\item Users and businesses are modeled with latent factor vectors computed over the training set using \textit{SVD++}. 
	\item Convolutional neural network is trained to predict businesses latent factors modeled in step 1 using the pre-processed review descriptions in the training set.
			Following parameters were used in the process of training the network: \textit{learning rate = 0.001}, stochastic gradient descent (\textit{SGD}) as optimization method and \textit{batch size} (number of instances forward-propagated before applying update of the network weights) was set to 64.
			For the purpose to evaluate the performance of the learned neural network model after each epoch, 10\% of the reviews from the training set are used as a validation set.
			Lowest RMSE is achieved in 23-rd epoch as shown on fig. \ref{fig:conv_net_train_test_error}. Hence, network model computed in the 23-rd epoch is used to obtain business latent factors.
	\item As an error function on the output layer we used Root Mean Squared Error (RMSE) between latent factors predicted by the network and the latent factors learned with matrix factorization.
	\end{enumerate}  
	
	\subsection{Results and discussion}
		
	We compare our method with baseline approach for overcoming item-cold start problem. The baseline approach estimates latent factor vectors of businesses in the test set by assigning random values for the latent factors. Two different approaches for random initialization are used. As another benchmark, we compare the results of the proposed solution with the results of the predictions obtained only with matrix factorization algorithm (SVD++) under the assumption that when evaluating on each of the test sets separately, the test set is also part of the training set (latent factor vectors are obtained from usage data). This is an upper bound to what is achievable when predicting the latent factors using business reviews. For all comparisons, Root Mean Squared Error (RMSE) is used as an evaluation measure on the predicted business ratings for all the users.
	
	The obtained results from the comparison are given in Table \ref{tbl:rmse_column-min-max_random_initialization}. It shows the results for the RMSE scores achieved when the latent factor vectors are randomly generated (\textit{Random 1} and \textit{Random 2}), latent factors are generated by using deep convolutional neural network (\textit{Proposed app.}) and when they are learned from usage data using \textit{SVD++}. \textit{Random 1} initialization assigns random values for the latent factors in the test set using uniform distribution on the interval between the lowest and the highest computed latent factor values from the training set. \textit{Random 2} initialization assigns random values from the interval between the lowest and the highest computed factor values, but only for the corresponding factor. If each business latent factors represent separate row in one matrix, then this kind of initialization assigns a value from the interval [\textit{min}, \textit{max}] of its corresponding column to each latent factor. The estimations of the RMSE scores for the both random initializations are performed on 100 runs and the average RMSE score and the variance are reported in the Table \ref{tbl:rmse_column-min-max_random_initialization}.
		
	\begin{table}[]
			\centering
			\caption{Results for RMSE scores achieved when the latent factor vectors are randomized (\textit{Random 1} and \textit{Random 2}), latent factors are generated by using deep convolutional neural network (\textit{Proposed app.}) and when they are learned from usage data using \textit{SVD++} (the upper bound scores)}
			\label{tbl:rmse_column-min-max_random_initialization}
			\begin{tabular}{|c|c|c|c|}
				\hline
				& \textbf{Test set 1} & \textbf{Test set 2} & \textbf{Test set 1 + Test set 2}            \\  \hline
				Random 1 &	 2.2070$\pm$0.0005          & 1.8309$\pm$9e-05        		& 1.8753$\pm$7e-05     \\  \hline
				Random 2 &	 1.8990$\pm$0.0002          & 1.5659$\pm$4e-05        		& 1.6054$\pm$4e-05     \\  \hline
				Proposed app. &	1.4094               		& 1.1300                		& 1.1663               \\  \hline
				SVD++ &		0.6850                 		& 0.5676               		& 0.5877                 \\  \hline
			\end{tabular}
		\end{table}

		According to the results shown in Table \ref{tbl:rmse_column-min-max_random_initialization}, the proposed approach outperforms the baseline approach (\textit{Random 1}) improving the RMSE score for \textbf{0.7976} for the test set 1, for \textbf{0.7009} for the test set 2 and for \textbf{0.7090} for the test set combined from the test set 1 and test set 2.
		
		The proposed approach also shows better results in comparison to the baseline approach that use \textit{Random 2} initialization method with improvement on the RMSE score for \textbf{0.4896} on the test set 1, for \textbf{0.4359} for the test set 2 and for \textbf{0.4391} for the test set combined from the test set 1 and test set 2.
	
	It has been experimentally shown \cite{Koren:2008:FMN:1401890.1401944} that a small improvement of the RMSE score has high influence on the quality of the top-K recommendations offered to a user. Therefore, from the improvements of the RMSE score that our model shows in comparison to the baseline estimate scores, we can conclude that our proposed algorithm has a significant impact in successfully overcoming the item-cold start problem. In this context, our model achieves highest improvements on the test set 1. This result is very important because test set 1 contains only the least reviewed businesses in our dataset, which best represent item-cold start problem. 
	\par
	There is a large gap between our result and the theoretical maximum (results obtained by using \textit{SVD++} on the whole dataset), but this is to be expected. Many aspects of the businesses that could influence the user preference cannot be extracted from the businesses reviews only. In our further work, we plan to use other information about the businesses such as description, popularity, type and location.
	
	\section{Conclusion}

In this paper, we investigated the use of deep learning for successfully addressing the item-cold start problem in recommendation systems. We use latent factor model for recommendation, and predict the latent factors from items descriptions using deep convolutional neural network when they cannot be obtained from usage data. 

We compare our approach to several baseline estimators using RMSE evaluation measure. The results have shown that the proposed approach significantly outperforms the baseline estimators showing a decrease of 0.7090 and 0.4391 of the RMSE score. Also, our results indicate that a lot of research in this domain could benefit significantly from using deep neural networks.

\section{Acknowledgments}
We would like to acknowledge the support of the European Commission through the project MAESTRA Learning from Massive, Incompletely annotated, and Structured Data (Grant number ICT-2013-612944). Also, this work was partially financed by the Faculty of Computer Science and Engineering at the Ss. Cyril and Methodius University. 
 
	\bibliographystyle{ieeetr}
	\bibliography{References}
	
\end{document}